\documentclass[11pt]{article}

\usepackage{epsfig}
\usepackage{amsmath}

\def\odk #1{(\ref{#1})}
\def\reff #1{Eq.~(\ref{#1})}
\def\kre#1{\hat{#1}^{\dagger}}
\def\vek#1{\boldsymbol{#1}}
\def\rr{\vek r}
\def\P#1{(\vek r_{#1},t_{#1})}

\begin{document}

\begin{center}
{\Large\bf Inequalities for Electron-Field Correlation Functions} \\[3mm]
{\large Tom\'a\v{s} Tyc}\\[2mm]
{\large \em Dept. of Theor. Physics, Masaryk University,  
         611 37 Brno, Czech Republic} \\
         email: tomtyc@physics.muni.cz\\[2mm]
June 2000\\[1mm]
{Published: Phys.~Rev.~A 62, 13803 (2000)}\\[4mm]
\end{center}
{\bf Abstract:} I show that there exists a class of inequalities between 
correlation functions of different orders of a chaotic electron field. 
These inequalities lead to the antibunching effect and are
a consequence of the fact that electrons are fermions -- indistinguishable
particles with antisymmetric states. The derivation of the inequalities 
is based on the known form of the correlation functions for the chaotic state
and on the properties of matrices and determinants.\\[7mm]

\section{Introduction}

In 1956 Hanbury Brown and Twiss observed a correlation of
photo-currents from two detectors aimed on the same star \cite{hbt}. They
explained this phenomenon using the classical electromagnetic theory of light. 
A more proper treatment of the problem shows that there must be a correlation
between the photons coming from the star.
Namely, photons are more likely to to arrive in groups (``bunches'') rather
than alone, which results in an enhanced shot noise with respect to randomly
arriving (Poisson) particles.
This phenomenon is called bunching and it is a typical behaviour of photons
emitted from thermal sources. It is caused by the fact that photons are not
distinguishable in principle and their quantum state is symmetrical with
respect to a permutation of two photons. 
In terms of the probability theory, bunching is expressed by the fact that
the probability of detecting two photons at the two detectors
shortly after one another is larger than the product of probabilities of
the two individual detections. 

For the case of electrons, a similar correlation has been predicted also in 1956
\cite{predpoved}. As an electron state is antisymmetrical with respect to
a permutation of two particles, electrons avoid coming in pairs which
results in a reduced shot noise. This phenomenon called antibunching
has been observed experimentally only recently \cite{harald}. In analogy to the
case of photons, antibunching is equivalent to the fact that the probability
of detecting two electrons at two detectors shortly after one another is less
than the product of probabilities of the two individual detections.

In this way, the typical behaviors of photons and electrons
can be expressed in terms of certain inequalities between the detection
probabilities. There is a question whether there is maybe a whole class of
inequalities between some physical quantities that would describe the bosonic
or fermionic behaviour of photons and electrons, respectively. We will show
that such inequalities exist, at least for electron chaotic states. To do
this, we first introduce correlation functions of the electron field.

\section{Correlation functions}

Suppose we have an electron field with the density operator
$\hat\rho$ and an electron detector with the quantum efficiency $\eta$ and the
cross-section $S$ localized at the point $\rr$, that is able to detect single
electrons. The probability of detecting an electron at the detector during
a short time interval $\langle t,t+\Delta t\rangle$ can be then expressed as
\begin{equation}
 P(\rr,t,\Delta t)=G^{(1)}(\rr,t)\,\eta S\Delta t,
\end{equation}
where $G^{(1)}(\rr,t)$ is the so-called one-electron correlation function
defined by the relation
\begin{equation}
  G^{(1)}(\rr,t)={\rm Tr}
  \bigl\{\hat\rho\kre\phi(\rr,t)\hat\phi(\rr,t)\bigr\},
\label{corr_1}\end{equation}
$\hat\phi(\rr,t)$ and $\kre\phi(\rr,t)$ being the flux
annihilation and creation operators of the electron at the space-time point
$(\rr,t)$ (see \cite{silverman}).

Now, suppose we have $k$ detectors at different points $\rr_1,\rr_2,\dots\rr_k$
and inquire what is the probability that we detect an electron at the first
detector during the time interval $\langle t_1,t_1+\Delta t\rangle$, another
electron at the second detector during the time interval
$\langle t_2,t_2+\Delta t\rangle$, etc., and the $k$th electron at the last
detector during the time interval $\langle t_k,t_k+\Delta t\rangle$.
This probability is now equal to 
\begin{equation}
 P(\rr_1,\dots,\rr_k,t_1,\dots,t_k,\Delta t)
  =G^{(k)}(\rr_1,\dots,\rr_k,t_1,\dots,t_k)\, (\eta S\Delta t)^k
\label{umera}\end{equation}
with the $k$-electron correlation function
\begin{equation}
  G_{1,2,\dots,k}^{(k)}\equiv G^{(k)}(\rr_1,\dots,\rr_k,t_1,\dots,t_k)={\rm Tr}
 \bigl\{\hat\rho\kre\phi(\rr_1,t_1)\cdots\kre\phi(\rr_k,t_k)
  \hat\phi(\rr_k,t_k)\cdots\hat\phi(\rr_1,t_1)\bigr\}.
\label{corr_k}\end{equation}

In principle, it is possible to evaluate the correlation functions for any
electron field according to \reff{corr_k}. However, the calculation can be
sometimes very difficult and correlation functions are nowadays known for
relatively few electron states \cite{Saito,tyc,klubka,ja}.
We will concentrate on an electron chaotic state in the following that is quite
explored and the explicit form of correlation functions is known for it.

The chaotic state is a generalization of a thermal state and it is believed
to be produced by the most coherent electron source nowadays available,
the field-emission gun \cite{ja,fieldemis}.
It is defined to have a maximum entropy if certain parameters
(the mean number of particles and the energy spectrum) are fixed
at given values. In other words, for these fixed parameters the chaotic
electron field is as random as possible. One of the interesting properties of
this state is that if there is some correlation 
in the chaotic field, it must have its origin in the indistinguishableness of
particles, i.e., in the Pauli principle. Indeed, distinguishable chaotic
particles would come to a detector completely uncorrelated which means 
that any joint detection probability would factorize into a product
of the individual detection probabilities, i.e., it would imply
$G^{(k)}_{1,2,\dots k}=G^{(1)}_1G^{(1)}_2\cdots G^{(1)}_k$. 
In this way, any aberration from this equation has its origin in the fermionic
nature of electrons.

According to \cite{Saito,ja}, the correlation function of a spin-polarized
chaotic state has the form of the determinant
\begin{equation}
   G_{1,2,\dots,k}^{(k)}=\left|\begin{array}{cccc}
    \Gamma_{11} & \Gamma_{12} & \dots & \Gamma_{1k} \\
    \Gamma_{21}&\Gamma_{22}&\dots&\Gamma_{2k}\\
    \vdots&\vdots& &\vdots\\
    \Gamma_{k1}&\Gamma_{k2}&\dots&\Gamma_{kk}\\
    \end{array}\right|,
\label{det}\end{equation}
where $\Gamma_{ij}
       ={\rm Tr}\bigl\{\hat\rho\kre\phi(\rr_i,t_i)\hat\phi(\rr_j,t_j)\bigr\}$
is the cross-correlation function of the electron field at the space-time
points $(\rr_i,t_i)$ and $(\rr_j,t_j)$.

It is useful to introduce the complex degree of coherence by the relation
\begin{equation}
   \gamma_{ij}=\frac{\Gamma_{ij}}{\sqrt{\Gamma_{ii}\Gamma_{jj}}}
\label{gamma}\end{equation}
(we suppose that $\Gamma_{ii}\not=0$ for all $i$; the opposite case is not
very interesting since some of the detectors are then not illuminated by
electrons at all).
An analogous physical quantity has been known in optics for a~long time
that expresses the~mutual coherence of the~electromagnetic field at two
space-time points\footnote{
One usually speaks about coherence of light but not about mutual coherence.
The coherence expresses the ability of light to interfere. 
In a similar way, if there is a mutual coherence of the electromagnetic
field at two points, there would occur interference if we brought the light
from these two points together.}.
Similarly, $\gamma_{ij}$ expresses the~mutual coherence of the~electron
field at the~space-time points  $\P i$ and $\P j$ and contains information
about both the~temporal and spatial
coherence of the~field.
As we will see later, the~matrix $\Gamma^{(k)}$ composed of the cross
correlation functions $\Gamma_{ij}$ is either positive-definite or
positive-semidefinite, from which it follows that
$\Gamma_{ij}\Gamma_{ji}\leq\Gamma_{ii}\Gamma_{jj}$ and
$|\gamma_{ij}|\leq1$ for all $i,j$. 
The~case $|\gamma_{ij}|=1$ corresponds to the~complete mutual coherence of 
the~electron field at the~points $\P i,\P j$, while $|\gamma_{ij}|=0$
corresponds to the~complete incoherence.
Thus for $|\gamma_{ij}|>0$, some properties of the~electron field at
the~point $\vek r_j$ at the~time $t_j$ can be determined from
the~knowledge of the~electron field at the~point $\vek r_i$ at the~time $t_i$.
On the~other hand, if $|\gamma_{ij}|=0$,
even if the~properties of the~field at the~point $\vek r_i$
at the~time $t_i$ are known completely, we cannot say anything
about the~field at the~point $\vek r_j$ at the~time $t_j$.

Using the properties of determinants and the fact that $\Gamma_{ii}=G^{(1)}_i$,
it is possible to re-write \reff{det} in terms of the $\gamma$'s:
\begin{equation}
G^{(k)}_{1,2,\dots, k}
  =G^{(1)}_1G^{(1)}_2\cdots G^{(1)}_k\left|\begin{array}{cccc}
   1 & \gamma_{12} & \dots & \gamma_{1k} \\
   \gamma_{21}&1&\dots&\gamma_{2k}\\
   \vdots&\vdots& &\vdots\\
   \gamma_{k1}&\gamma_{k2}&\dots&1\\
   \end{array}\right|.
\label{DETT}\end{equation}

\section{Inequality between one- and two-electron correlation functions}
\begin{figure}[htb]
\begin{center}
\mbox{\epsfxsize=8cm\epsfbox{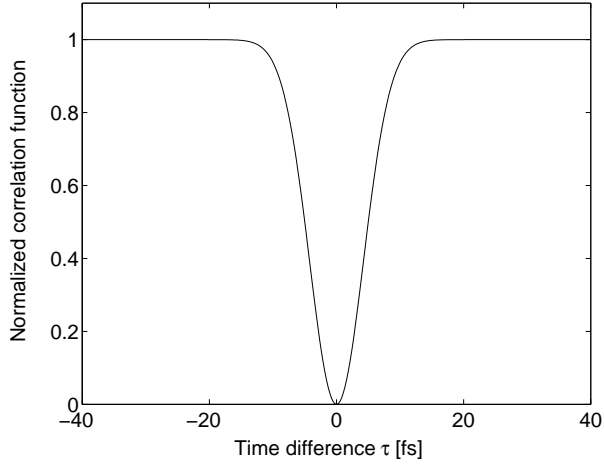}}
\caption{\small The normalized two-electron correlation function
$G_{1,2}^{(2)}/[G_1^{(1)}G_2^{(1)}]=1-|\gamma_{12}|^2$ for $\rr_1=\rr_2$
as a function of the time difference $\tau=t_2-t_1$ for spin-polarized
quasi-monochromatic electrons with the Gaussian energy spectrum. The width of
the peak is approximately equal to the coherence time of the electrons. The
probability that two electrons come after one another within the coherence
time is less than the probability that they come longer after one another,
which is called antibunching.}
\label{anti}
\end{center}
\end{figure}

We will first investigate the two-electron correlation function. According
to \reff{DETT} it follows that
\begin{equation}
   G_{1,2}^{(2)}=G_1^{(1)}G_2^{(1)}\left(1-\gamma_{12}\gamma_{21}\right)
   =G_1^{(1)}G_2^{(1)}\left(1-|\gamma_{12}|^2\right).
\label{2}\end{equation}
Here we used the fact that $\gamma_{12}=\gamma^*_{21}$ that will be proved
later. The equation \odk{2} shows that 
\begin{equation}
G_{1,2}^{(2)} \le G_1^{(1)}G_2^{(1)},
\label{nerov}\end{equation}
so the joint detection probability is less than or equal to the product of the
individual
detection probabilities. It means that one is not likely to detect two electrons
at the space-time points where the electron field is mutually coherent.
In usual electron fields, this happens
if the spatial separation of the two points $\P1$ and $\P2$ is not larger
than the coherence length $l_c$ of the electrons and if the time difference
$t_2-t_1$ is not larger than the coherence time $T_c$ \footnote{
Strictly speaking, in quasi-monochromatic fields $\gamma_{12}\not=0$ holds
if $\rr_2-\rr_1\approx\vek v(t_2-t_1)$, where $\vek v$ is the group velocity
of the electrons}. From this follows
that a detection of two electrons at the same detector
with a time separation less than $T_c$ is not likely because the term
$1-|\gamma_{12}|^2$ is then small. On the other hand, the detection
probability of two electrons with a time separation much more than $T_c$
(when $\gamma_{12}$ is already equal to zero) is simply equal to the product
of the individual detection probabilities and there is therefore  no
correlation.
So it seems that at the typical time scale of $T_c$, the electrons avoid
coming in pairs (or groups) to a detector and prefer coming alone.
This effect is called antibunching (see Fig.~\ref{anti}).
Thus, we can say that antibunching is a consequence
of the fact that the probability of detecting two electrons
at two detectors shortly after one another is less than the product of
the probabilities of the two individual detections, or more generally,
that it is a consequence of the inequality \odk{nerov}.

In the~extreme case when $|\gamma_{12}|=1$, the two-electron correlation
function
turns into zero. Then no two electrons can be found at the~space-time points
$\P 1$ and $\P 2$ simultaneously. This reminds one of the~Pauli principle:
the~latter prohibits two electrons to be in the~same quantum state,
while \reff{2} prohibits two electrons to be at the space-time points
$\P 1$ and $\P 2$ where the electron field is mutually completely
coherent.

The inequality \odk{nerov} holds between the one- and two-electron correlation
functions of a chaotic electron field. Now, the question is whether it would be
possible to find a similar inequality also between correlation functions of
higher orders. The answer is yes.
A possible generalization of \odk{nerov} that comes to mind is 
$G^{(k)}_{1,2,\dots,k}\leq G^{(1)}_{1}\,G^{(1)}_{2}\cdots G^{(1)}_{k}$.
If this should hold, then the determinant in \reff{DETT} would have to be less
than or equal to unity. In the following we will show that it is indeed so
by using the well-known properties of matrices and determinants. Moreover, we
will prove an even more general inequality between the correlation functions of
different orders.

\section{General inequality between correlation functions}
First we note that the matrix composed of the~cross-correlation functions
 \begin{equation}
   \Gamma^{(k)}=\left(\begin{array}{cccc}
   \Gamma_{11} & \Gamma_{12} & \dots & \Gamma_{1k} \\
   \Gamma_{21}&\Gamma_{22}&\dots&\Gamma_{2k}\\
   \vdots&\vdots& &\vdots\\
   \Gamma_{k1}&\Gamma_{k2}&\dots&\Gamma_{kk}\\
 \end{array}\right)
 \label{matrix}\end{equation}
is Hermitian and either positive-definite or positive-semidefinite.
The~hermiticity of $\Gamma^{(k)}$ follows simply from the Hermiticity of the
density operator $\hat\rho$ and from the invariance of the
trace under a commutation of operators:
\begin{equation}
\Gamma_{ij}={\rm Tr}\bigl\{\hat\rho\kre\phi_i\hat\phi_j\bigr\}
           =\bigl[{\rm Tr}\bigl\{\kre\phi_j\hat\phi_i\kre\rho\bigr\}\bigr]^*
           =\bigl[{\rm Tr}\bigl\{\hat\rho\kre\phi_j\hat\phi_i\bigr\}\bigr]^*
           =\Gamma_{ji}^*.
\label{dukaz}\end{equation}
Of course, from \reff{dukaz} it follows also that $\gamma_{ij}=\gamma^*_{ji}$,
i.e., the matrix $\gamma^{(k)}$ composed of the complex degrees of coherence
is also Hermitian.

The~second property can be proved in a similar way as
an~analogous statement in the quantum optics (see \cite{Mandel}, p. 585).
Let $\hat O$ be the~operator defined as
\begin{equation}
  \hat O=\sum_{i=1}^k {\lambda_i}\,\hat\phi_i,
\label{O}\end{equation}
where $\lambda_1,\dots,\lambda_k$ are arbitrary complex numbers.
It holds
\begin{equation} 
 {\rm Tr}\{\hat\rho\,\kre O\hat O\}
   =\sum_{i,j=1}^k \lambda_i^*\lambda_j
       {\rm Tr}\bigl\{\hat\rho\kre\phi_i\hat\phi_j\bigr\}
   =\sum_{i,j=1}^k \lambda_i^*\lambda_j\,\Gamma_{ij}.
\label{definit}\end{equation}
At the~same time, ${\rm Tr}\{\hat\rho\,\kre O\hat O\}$
is a~non-negative number. As the~right-hand side of \reff{definit} is
a quadratic form in the~$\lambda$'s with the coefficients $\Gamma_{ij}$,
the~matrix $\Gamma^{(k)}$ 
must be either positive-definite or positive-semidefinite.
A similar statement can be proved also for the matrix $\gamma^{(k)}$ using
$\hat O=\sum_{i=1}^k {\lambda_i}\,(\Gamma_{ii}\Gamma_{jj})^{-1/2}\,\hat\phi_i$.

As we will see in the following, from the Hermiticity and definiteness of the
matrices $\Gamma^{(k)}$ and $\gamma^{(k)}$ the desired inequality follow
directly. However, first it will be necessary to prove the following lemma:

\noindent{\bf
Lemma:}\\
The~determinant of any positive-definite or positive-semidefinite
Hermitian matrix $A=(A_{ij})$ with nonzero diagonal elements
cannot exceed the~product of the~diagonal elements of $A$,
i.e., ${\rm det}(A)\leq A_{11}A_{22}\cdots A_{kk}$, and the
equality takes place if and only if $A$ is diagonal.

\noindent{\bf
Proof:}\\
As all the diagonal elements $A_{ii}$ of the matrix $A$ are positive,
we can define the~matrix $a=(a_{ij})$ with
elements $a_{ij}=A_{ij}/\sqrt{A_{ii}A_{jj}}$ (in analogy with defining the
matrix $\gamma^{(k)}$ with the help of $\Gamma^{(k)}$).
Thanks to the~hermiticity of the~matrix $a$, it is possible to transform it
into the~diagonal form with a~unitary transformation, i.e., there exists
a~unitary matrix $U$ for which the
matrix $b=UaU^\dagger$ is diagonal. This transformation changes neither the
determinant nor the~trace of the~matrix because it is a~unitary transformation.
If we denote the~diagonal elements of the~matrix $b$ as $b_i$,
then ${\rm Tr}(a)={\rm Tr}(b)=\sum_{i=1}^k b_i$ and
${\rm det}(a)= {\rm det}(b)=\prod_{i=1}^k b_i$
evidently hold.  At the~same time,
${\rm Tr}(a)=k$ holds due to the fact that $a_{ii}=1$ for all $i$.
To find out what is the~maximal possible value of ${\rm det}(a)$, we will
use now the~inequality between the~arithmetical and geometrical averages.
The~arithmetical average of the~numbers $b_i$ is $\alpha=\sum_{i=1}^kb_i/k=1$
and their geometrical average is $\beta=\sqrt[k]{\prod_{i=1}^k b_i}$.
As the~numbers $b_i$ are non-negative, the~inequality $\beta\leq\alpha$ holds,
from which it then follows that
${\rm det}(a)= \prod_{i=1}^k b_i\leq1$.
As is known, the~equality $\beta=\alpha$ takes place if and only if
$b_1=b_2=\ldots=b_k$. In this case the~matrix $b$ is the~unit
matrix, from which it follows that $a$ is also the~unit matrix and
$a_{ij}=\delta(i,j)$. Thus, ${\rm det}(a)\le 1$ holds and the~equality
takes place only when all the~non-diagonal elements of the~matrix $a$ vanish. 
Expressing the~determinant of the original matrix $A$ with the~help of
${\rm det}(a)$ as ${\rm det}(A)=A_{11}A_{22}\cdots A_{kk}\,{\rm det}(a)$,
we get from the~inequality ${\rm det}(a)\le 1$ that
\begin{equation}
  {\rm det}(A)\leq A_{11}A_{22}\cdots A_{kk}.
\label{nerovk}\end{equation}
Moreover, $A$ is diagonal if and only if $a$ is diagonal. Therefore the
equality in \odk{nerovk} takes place if and only if the~matrix $A$ is diagonal.

If we identify the matrix $a$ with $\gamma^{(k)}$, then from \reff{DETT} and the
proof above it follows immediately that
\begin{equation} 
  G^{(k)}_{1,2,\dots,k}
  \leq G^{(1)}_{1}\,G^{(1)}_{2}\cdots G^{(1)}_{k}.
\label{stara}\end{equation}
This is a generalization of the inequality \odk{2} for a correlation function
of arbitrary order. An even stronger generalization would be evidently
\begin{equation} 
  G^{(k)}_{1,2,\dots,k}
        \leq G^{(l)}_{1,2,\dots,l}\,G^{(k-l)}_{l+1,l+2,\dots,k}.
\label{ineq2}\end{equation}
As we will see now, this inequality indeed holds.

\section{Proof of the inequality (16)} \label{secdukaz}

First we will define a matrix $\Gamma'$ of the~type $k/k$ in the~following
block form:
\begin{equation}
 \Gamma' =\left(\begin{array}{cc}
  \Gamma^{(l)} & 0  \\
  0 &\Gamma^{(m)}\\
  \end{array}\right).
\end{equation}
Here 0 stands for the~zero matrices of the~type $l/m$ or $m/l$
(we have denoted $m=k-l$) and
$\Gamma^{(l)},\Gamma^{(m)}$ are the~matrices of the~type $l/l$ and $m/m$,
respectively, corresponding to the~correlation functions
$G^{(l)}_{1,\dots,l}$ and $G^{(m)}_{l+1,\dots,k}$:
\begin{equation}
   \Gamma^{(l)}=\left(\begin{array}{cccc}
   \Gamma_{1,1}& \dots & \Gamma_{1,l} \\
   \vdots& &\vdots\\
   \Gamma_{l,1}&\dots&\Gamma_{l,l}\\
   \end{array}\right), \qquad
   \Gamma^{(m)}=\left(\begin{array}{cccc}
   \Gamma_{l+1,l+1} &\dots & \Gamma_{l+1,k} \\
   \vdots& &\vdots\\
   \Gamma_{k,l+1}&\dots&\Gamma_{k,k}\\
   \end{array}\right)
\end{equation}
Due to \reff{det} and the~block form of $\Gamma$ it holds
\begin{equation} 
  G^{(k)}_{1,\dots,k}={\rm det}(\Gamma),\quad
  G^{(l)}_{1,\dots,l}\,G^{(m)}_{l+1,\dots,k}
   ={\rm det}(\Gamma^{(l)})\,{\rm det}(\Gamma^{(m)})={\rm det}(\Gamma').
\label{corrdet}\end{equation}
Now, we know that the~matrix $\Gamma\equiv\Gamma^{(k)}$ is either
positive-definite
or positive-semidefinite. In the~latter case, the~inequality \odk{ineq2} is
satisfied trivially because then ${\rm det}(\Gamma)=0$ and
${\rm det}(\Gamma^{(l)}),{\rm det}(\Gamma^{(m)})$ are both non-negative due to
their definiteness. Therefore in the~following we will discuss the~case when
$\Gamma$ is positive-definite.

As the~matrices $\Gamma^{(l)}$ and $\Gamma^{(m)}$ are Hermitian,
it is possible to transform each of them
into the~diagonal form with a~unitary transformation.
Let $U^{(l)}$ and $U^{(m)}$ denote the~corresponding
unitary transformational matrices, so that the~matrices
$D^{(l)}=U^{(l)}\Gamma^{(l)} U^{(l)\dagger}$ and
$D^{(m)}=U^{(m)}\Gamma^{(m)} U^{(m)\dagger}$ are both diagonal.
Then evidently the~unitary matrix
\begin{equation}
 U=\left(\begin{array}{cc}
  U^{(l)} & 0  \\
  0 &U^{(m)}\\
  \end{array}\right)
\end{equation}
transforms the~matrix $\Gamma'$ into the~diagonal form, so that
$D'=U\Gamma'U^\dagger$ is diagonal.
Let $D$ denote the~matrix obtained from $\Gamma$ by the~same unitary
transformation, i.e., $D=U\Gamma U^\dagger$. Thanks to the~block form of
the~matrix $U$, the~matrix $D$ has the~form
\begin{equation}
 D=\left(\begin{array}{cc}
  D^{(l)} &  D^{(lm)}  \\
  D^{(ml)} &D^{(m)}\\
  \end{array}\right),
\label{decka}\end{equation}
where $D^{(lm)}$ and $D^{(ml)}$ are some mutually Hermite-conjugate matrices
of the~type $l/m$ and $m/l$, respectively.
Applying now Lemma to the~matrix $D$
(we can do that because $D$ is positive-definite and  Hermitian; the~latter
follows from the~unitarity of the~matrix $U$),
we see that ${\rm det}(D)\leq{\rm det}(D')$
because the~diagonal elements of the~matrices $D$ and $D'$ are identical
and $D'$ is diagonal.
Combining this with the~equations that hold
due to the~unitarity of the~matrix $U$,
\begin{equation} 
 {\rm det}(D)={\rm det}(\Gamma), \qquad {\rm det}(D')
  ={\rm det}(\Gamma')={\rm det}(\Gamma^{(l)})\,{\rm det}(\Gamma^{(m)}),
\label{determinanty}\end{equation}
and with \reff{corrdet}, we finally obtain the~inequality \odk{ineq2}.
Now, the inequality  ${\rm det}(D)\leq{\rm det}(D')$ changes into equality
if and only if the~matrix $D$ is diagonal, i.e., if $D^{(lm)}$
and $D^{(ml)}$ are the~zero matrices. Then, again due to the~block form of the
transformation matrix $U$, also the~matrices  
\begin{equation}
   \Gamma^{(lm)}=\left(\begin{array}{ccc}
   \Gamma_{1,l+1}& \dots & \Gamma_{1,k} \\
   \vdots& &\vdots\\
   \Gamma_{l,l+1}&\dots&\Gamma_{l,k}\\
   \end{array}\right), \qquad 
   \Gamma^{(ml)}=\left(\begin{array}{cccc}
   \Gamma_{l+1,1} &\dots & \Gamma_{l+1,l} \\
   \vdots& &\vdots\\
   \Gamma_{k,1}&\dots&\Gamma_{k,l}\\
   \end{array}\right)
\end{equation}
are the~zero matrices.
Thus we can conclude that the~inequality \odk{ineq2} holds and it changes into
equality if and only if all the~cross-correlation functions $\Gamma_{i,j}$
vanish for $i=1,\dots,l$ and $j=l+1,\dots,k$.

\section{Fermionic nature of electron correlations}

\begin{figure}[htb]
\begin{center}
\mbox{\epsfxsize=7.5cm\epsfbox{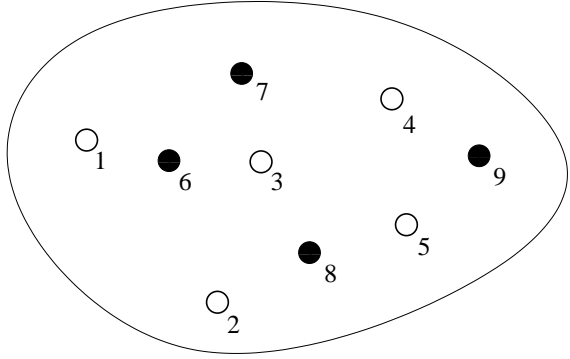}}
\caption{\small An example of inequality \odk{ineq2} for $k=9$, $l=5$.
The correlation function corresponding to all the points is less
than or equal to the product of correlation functions corresponding to the
``white'' and ``black'' points. The equality takes place if and only if
$\gamma_{ij}=0$ for each pair containing one white and one black point.}
\label{body}
\end{center}
\end{figure}

Let us see what the inequality \odk{ineq2} that we just proved really means.
We denote the detection of $l$ electrons at the~space-time points
$\P1,\dots,\P{l}$ as event A and the the detection of $k-l$ electrons at
the~points $\P{l+1},\dots,\P k$ as event B. Then the inequality
\odk{ineq2} says that the probability that both events A and B happen is less
than or equal to the product of probabilities of events A and B (see
Fig.~\ref{body}). In this way, the inequality
\odk{ineq2} generalizes the inequality \odk{nerov} also to multiple electron
detection processes.

We have seen that the inequality \odk{nerov} leads to the antibunching effect
as a consequence of the indistinguishableness of electrons.
Similarly, the inequality \odk{ineq2} reflects the same principle
for more complicated detection processes and leads to more general
correlations in electron fields. 
It is a~fundamental statement that expresses the~fermionic behaviour of
electrons in a~very compact way. 

As follows from Sec.~\ref{secdukaz}, the~case of equality in \odk{ineq2} corresponds to
the situation when $\Gamma_{ij}=0$ (and hence $\gamma_{ij}=0$)
for all $i=1,\dots,l$ and $j=l+1,\dots,k$. Then the electron field at any point
of the~first set of points $S_l=\{\P i|\,i=1,\dots,l\}$ is incoherent with
the~field at any point of the~second set $S_{m}=\{\P j|\,j=l+1,\dots,k\}$.
The~equality in \odk{ineq2} is then very reasonable: if the~fields at
the~points corresponding to the~both sets $S_l,S_m$
are mutually completely incoherent, the~detections at the points of the
two sets are mutually independent and therefore total detection probability
factorizes into the~product of the~detection probabilities corresponding
to the~individual sets.

Of course, the~inequality \odk{ineq2} can be applied repeatedly and the~points
$\P1,\dots,\P{k}$ can be interchanged arbitrarily to obtain a whole class of
inequalities. We will write just an~example for illustration:
\begin{equation} 
  G^{(7)}_{1,2,3,4,5,6,7}\leq G^{(2)}_{1,2}\,G^{(3)}_{3,5,7}\,G^{(1)}_{4}
                              \,G^{(1)}_{6}.
\end{equation}

\section{Conclusion}

We have proved a relatively simple inequality between
correlation functions of different orders for chaotic electrons.
As any correlation in a chaotic electron field originates from the
fermionic character of the electrons, the inequality \odk{ineq2}
is a direct consequence of the Pauli principle. It demonstrates the
aversion of the electrons to staying and coming to a detector in groups. 
The inequality \odk{ineq2} determines a set of conditions that must
be fulfilled on the hierarchy of the chaotic correlation functions.
We must point out that the inequalities \odk{ineq2} do not hold
for all electron fields. There are electron states that show even bunching
instead of antibunching \cite{silverman3,silverman4}. However, these states are
quite rare and the chaotic state remains the most important and wide-spread
state in electron beams.

>From the experimental point of view, the observation of correlations 
is limited especially by an extremely short coherence time of
available electron beams.  The coherence time is related to the energy
bandwidth of the beam by the relation $T_c\approx h/\Delta E$, which yields
$T_c\approx 2\times10^{-14}\,\rm s$ for a typical fiels-emission beam for which
$\Delta E\approx0.2\,\rm eV$. The measurement of correlations with
such a characteristic time requires very fast detectors and coincidence
electronics and even under optimum conditions the experimental resolution 
time exceeds the coherence time by three orders of magnitude. The
signal-to-noise ratio is therefore very small and it is not surprising that
a two-electron correlation was observed in the last year only. Observation of
higher-order correlations would require a more complicated experimental setup
and I believe that it will not be possible until electron sources with a much
longer resolution time become available. On the other hand, if the resolution
and coherence time became comparable, the highest order of observable
correlations would be limited by the fidelity of the coincidence electronics.
Thus we must conclude that the inequality \odk{nerov} is the only
candidate for an experimental verification from the whole class of inequalities
\odk{ineq2} at the present time.

\section*{Acknowledgments}
I would like to thank professor M.~Lenc for helpful discussions. This work
was supported by the Czech Ministry of Education, contract No. 144310006.

\end{document}